\documentclass[showkeys,showpacs,prd]{revtex4}
\usepackage{amsmath}
\usepackage{amssymb}
\usepackage{graphicx}
\usepackage{color}
\usepackage{enumitem}

\begin{document}

\title{
Nonperturbative quantization: ideas, perspectives, and applications 
}

\author{
Vladimir Dzhunushaliev$^{1,2}$
\footnote{Email: v.dzhunushaliev@gmail.com}
}
\affiliation{$^1$
Dept. Theor. and Nucl. Phys., KazNU, Almaty, 050040, Kazakhstan; \\
$^2$ IETP, Al-Farabi KazNU, Almaty, 050040, Kazakhstan \\
}

\begin{abstract}
The procedure of nonperturbative quantization \`{a} la Heisenberg is considered.
A few applications, features, perspectives, problems, and so on are considered. The comparison with turbulence modeling is performed.
\end{abstract}


\maketitle

\tableofcontents

\section{Introduction}
\label{intr}

At the present time
the problem of quantization is mathematically well defined only for free fields.
Despite its successes, a perturbative (P) quantization has its own problems: (a) renormalization is not mathematically well-defined procedure (in this regard Feynman said that it is just sweeping the garbage under the rug); (b) series constructed from Feynman diagrams are only asymptotic series, which, strictly speaking, are not convergent series.

 Perturbative quantum field theory is based on the notion of interacting point particles.
 In this case a mathematical tool for describing the interaction of particles are Feynman diagrams. They describe point particles that interact at the vertices (points in spacetime) and move between these interaction acts as free particles. This description can be used only for small coupling constants and leads to such a problem as the problem of renormalization.  From the physical point of view this problem is explained as follows: the propagator of free particles is a distribution, and in describing loops the product of distributions is appeared which is badly defined from the mathematical point of view. Fortunately, the renormalization procedure of physical quantities leads to finite results. R. Feynman, one of founders of perturbative calculations, called this procedure ``sweeping the garbage under the rug.''

The renormalization procedure divides all theories into two classes: renormalizable and non-renormalizable ones. In this regard, we note the following: When we work with Feynman diagrams, physically it means that we use the concept of a particle, and vice versa, if we introduce the notion of quantum particles we have to use Feynman diagrams. That is, we have a one-to-one correspondence between Feynman diagrams and particles -- quanta. Here we must clearly understand that if we cannot use the concept of quantum particles, we cannot draw Feynman diagrams, that is true in reverse too: if we cannot draw Feynman diagrams, we cannot introduce the concept of quantum particles. This observation forces us to conclude that in nonperturbative (NP) quantum field theory (where one cannot use Feynman diagrams), we cannot introduce the concept of quanta.

Another interesting feature in NP quantum field theory is that there is no any renormalization procedure because the renormalization occurs only when we use Feynman diagrams. An important consequence of this observation is that the non-renormalizable theories can be quantized at the nonperturbative level. \textcolor{blue}{\textit{This means that non-renormalizability does not mean the non-quantalizability.}}

The results of P calculations are also in conflict with gravity. The problem is that in performing
 P calculations one inevitably deals with zero-point quantum vacuum fluctuations with nonzero energy
which, according to general relativity, are sources of a gravitational field. Computation of these corrections gives rise to the appearance of a
cosmological constant whose value exceeds its real value by many orders of magnitude.
From the physical point of view the appearance of this energy is related to
the fact that an interacting particle moves between interactions as a free particle,
and a free particle in quantum field theory is an oscillator having the nonzero lowest energy value.
This energy should give rise to the cosmological constant but, as mentioned above, the value calculated by using P calculations is not consistent with the
one
measured in astrophysical observations.

The first NP quantum calculations were made by W. Heisenberg, see Ref. \cite{heis}. He applied the NP quantization method for quantizing a nonlinear spinor field. The physical motivation for his calculations was the attempt to obtain the physical theory of an electron. Heisenberg's calculations allowed to evaluate some properties of an electron with some accuracy.

\section{Physical basis for the NP quantization}

Before we proceed to
a mathematical basis of NP quantization,
let us discuss the physical grounds of NP quantum physics.
We established in the Introduction that a NP quantum field is not a cloud of quanta. Before starting the mathematical description of quantization procedure,
it would be very useful to discuss the physical essence of a NP quantum field.
In our opinion a strongly interacting quantum field (that must be quantized using a NP quantization procedure) is similar to a turbulent fluid.
At any point we have a fluctuating field which is analogous to velocity, pressure, and so on for a turbulent fluid. Green's functions of NP quantum fields outside of a light cone are nonzero, in contrast to P~quantized fields. The reason is that for strongly interacting fields localized regular objects may exist. For example, such objects do exist in classical theories: non-Abelian monopoles and instantons, solitons, etc. It is possible to expect the existence of such objects in quantum theory of nonlinear  strongly interacting fields. In this case the correlation of quantum fields at two spatially separated points will be nonzero.

Now we can start the mathematical description of the procedure of NP quantization. First, we want to describe briefly this approach:
\begin{enumerate}[label=\bfseries Step \arabic*]
	\item
	\label{st1}	
	We write down field equations where all physical quantities
	are written with hats, i.e., for quantum field operators.
	\item
	\label{st2}	
	We write down the expectation value of these equations.
	\item
	\label{st3}	
	We see that the equations couple various Green's functions.
	For example, we have $\left\langle \phi \right\rangle$ and
	$\left\langle \phi^3 \right\rangle$ in one equation.
	\item
	\label{st4}	
	We multiply an initial field equation by some combination
	of field operators to get the equation for the desired Green's
	function. For example, multiplying the equation for
	$\hat \phi$ by $\hat \phi^2$, we obtain the equation with
	$\hat \phi^3$.
	\item
	\label{st5}	
	After the quantum averaging we get an equation for unknown
	Green's function. For example, we obtain the equation for
	$\left\langle \phi^3 \right\rangle$.
	\item
	\label{st6}	Again we see that the new equation contains a new
	Green's function. For example, it will be
	$\left\langle \phi^5 \right\rangle$.
	\item We repeat \ref{st4} - \ref{st6} once again, and so on {\it ad infinitum}.
\end{enumerate}
When we write down operator field equations it is not clear how we can work with the operator equations?
How we can describe the properties of this  operator? In the P quantization scheme, after the quantum averaging,
the operator equation for a free field gives us the equation for the propagator. This equation can be solved, since it is a linear equation.
As mentioned above,
in our case the situation is different: as a consequence of nonlinearity, we have to write an infinite set of equations
for all Green's functions. We think that the properties of the field operator are described by all Green's functions. This means that the operator equation is equivalent to a set of equations
for all possible Green's functions.  Simultaneously, all Green's functions give us full information about the quantum state $\left. \left. \right| Q \right\rangle$. Here we use
$\left. \left. \right| Q \right\rangle$ for the quantum averaging,
$
\left\langle \phi \right\rangle =
\left\langle Q \left|
	\phi 	
\right| Q \right\rangle
$.

Finally, we want to say that in practice we cannot solve directly the field operator equations. Instead of this,
we have to solve the infinite set of coupled equations for all Green's functions.
In doing so, we see the next problem:
how such infinite set of equations can be solved? We think that the only possibility is to cut off the
infinite set of equations to obtain a finite system of equations by using some physical assumptions about higher-order Green's functions.

Similar procedure is well known in modeling turbulence (see, for example, the textbook of Wilcox~\cite{Wilcox}). The situation there is as
follows (we follow Ref.~\cite{Wilcox} in this paragraph): One can write a statistically averaged version of the Navier-Stokes equation
(the Reynolds-averaged Navier-Stokes equation) for an averaged velocity. This equation contains six new unknown functions $\overline{\rho v_i v_j}$
(the Reynolds-stress tensor, where the overbar denotes statistical averaging). This means that our system is not yet closed. In quest of additional equations, we have to take moments of the Navier-Stokes equation. That is, we multiply the Navier-Stokes equation by a suitable quantity and statistically average the product. Using this procedure, we can derive a differential equation for the Reynolds-stress tensor. After such procedure we gained six new equations, one for each independent components of the Reynolds-stress tensor. However, we have also generated 22 new unknown functions: $\overline{\rho v_i v_j v_k}$,
$
\overline{\frac{\partial u_i}{\partial x_k}
\frac{\partial u_j}{\partial x_k}
}
$,
$
\overline{
u_i \frac{\partial p}{\partial x_j}
+ u_j \frac{\partial p}{\partial x_i}
}
$.
This situation illustrates the closure problem of turbulence modeling (let us note that we have the similar problem for a nonperturbative quantization).
Because of the nonlinearity of the Navier-Stokes equation, as we have higher and higher moments, we generate additional unknown functions
at each level. As written in Ref.~\cite{Wilcox}: ``The function of turbulence modeling is to derive approximations for the unknown
correlations in terms of flow properties  that are known so that a sufficient number of equations exists. In making such approximations, we close the system.''

Following this scheme, we can rephrase the last sentence as applied to a nonperturbative quantization:
\emph{The approximate  approach for a nonperturbative quantization being suggested here
is to derive approximations for unknown Green's functions using the properties of the quantum system under consideration
so that a sufficient number of equations exists. In making such approximations, we close an infinite set of equations for the Green's functions.}

After that point, we want to say that a NP quantized field has similar properties to a turbulent fluid:
(a) at every point we have fluctuating quantity(ies); (b) the NP quantum field(s) cannot be presented as a cloud of quanta; (c) the correlation between spatially separated points is nonzero.

\section{Mathematics of nonperturbative quantization}

In this section we follow Ref.~\cite{Dzhunushaliev:2013nea}. According to \ref{st1}, we write the operator field equation
\begin{equation}
	\frac{\partial }{\partial x^\mu} \left(
		\frac{\partial \hat{\mathcal L}}{\partial \hat \phi^A_{,\mu}}
	\right) -
	\frac{\partial \hat{\mathcal L}}{\partial \hat \phi^A} = 0,
\label{3-10}
\end{equation}
where $\hat{\mathcal L}$ is the operator of the Lagrangian density; $\hat \phi^A$ is the operator of the field; the index $A$ encodes all possible indices in one. This equation is the key equation for the NP quantization. According to \ref{st1}-\ref{st6}, we have to obtain equations for all Green's functions using this master equation. \textcolor{blue}{\emph{The main idea is that having all Green's functions we know the properties of given quantum state $\left. \left. \right| Q \right\rangle$
and field operators $\hat \phi^A$.}} It is evident that this receipt works for free fields.
Applying the aforementioned receipt for a free field, we obtain the propagator and all higher-order Green's functions  as polylinear combinations of propagators.

The first example: the operator Yang-Mills equation is
\begin{equation}
    \partial_\nu \widehat {F}^{a \mu\nu} =
    \hat J^{a \mu},
\label{3-60}
\end{equation}
where
$\hat F^B_{\mu \nu} = \partial_\mu \hat A^a_\nu -
\partial_\nu \hat A^b_\mu +
g f^{abc} \hat A^b_\mu \hat A^c_\nu$
is the field strength operator;
$\hat A^a_\mu$ is the gauge potential operator; $a,b,c = 1, \ldots ,8$ are the SU(3) color indices; $g$ is the coupling constant; $f^{abc}$ are the structure constants for the SU(3) gauge group; $\hat J^{a \mu}$ is the current operator.

The second example: in gravitation theory, we have the operator Einstein equation
\begin{equation}
	\hat R_{\mu \nu} - \frac{1}{2} \hat g_{\mu \nu} \hat R =
	\varkappa \hat T_{\mu \nu} ,
\label{3-20}
\end{equation}
where all geometrical quantities are defined in the usual manner from the corresponding operators:
\begin{eqnarray}
	\hat R_{\mu \nu} &=& \hat R^\rho_{\phantom{\rho} \mu \rho \nu},
\label{3-30}\\
	\hat R^\rho_{\phantom{\rho} \sigma \mu \nu} &=&
	\frac{\partial \hat \Gamma^\rho_{\phantom{\rho} \sigma \nu}}
	{\partial x^\mu} -
	\frac{\partial \hat \Gamma^\rho_{\phantom{\rho} \sigma \mu}}
	{\partial x^\nu} +
	\hat \Gamma^\rho_{\phantom{\rho} \tau \mu}
	\hat \Gamma^\tau_{\phantom{\tau} \sigma \nu} -
	\hat \Gamma^\rho_{\phantom{\rho} \tau \nu}
	\hat \Gamma^\tau_{\phantom{\tau} \sigma \mu} ,
\label{3-40}\\
	\hat \Gamma^\rho_{\phantom{\rho} \mu \nu} &=&
	\frac{1}{2} \hat g^{\rho \sigma} \left(
		\frac{\partial \hat g_{\mu \sigma}}{\partial x^\nu} +
		\frac{\partial \hat g_{\nu \sigma}}{\partial x^\mu} -
		\frac{\partial \hat g_{\mu \nu}}{\partial x^\sigma}
	\right),
\label{3-50}
\end{eqnarray}
and $\hat T_{\mu \nu}$ is the energy-momentum tensor operator.

Heisenberg's technique offers to use an infinite set of equations  for all Green's functions, which can be written as follows
\begin{eqnarray}
	\left\langle
	\text{ LHS Eq. \eqref{3-10}}
	\right\rangle &=& 0 ,
\label{3-70}\\
	\left\langle \hat \phi^A(x_1)
	\cdot \text{ LHS Eq. \eqref{3-10}}
	\right\rangle &=& 0 ,
\label{3-80}\\
	\left\langle \hat \phi^A(x_1) \hat \phi^B(x_2)
	\cdot \text{ LHS Eq. \eqref{3-10}}
	\right\rangle &=& 0 ,
\label{3-90}\\
	\cdots &=& 0	,
\label{3-100}\\
	\left\langle
	\text{ the product of $\hat \phi$ at different points $(x_1,
	\cdots , x_n)$} 	\cdot \text{ LHS Eq. \eqref{3-10}}
	\right\rangle &=& 0.
\label{3-110}
\end{eqnarray}

\section{Significant differences with respect to perturbative quantization}

It is believed within the NP quantization scheme for strongly interacting fields that an interaction happens always and everywhere. On the other hand,
the P quantization scheme for weakly interacting fields assumes that the fields are free almost always and everywhere, except vertices where the interactions happen. In this sense the P quantization procedure is contradictory: we use Lagrangian for free fields almost always and everywhere and only at the vertices we use full Lagrangian with interacting terms.

The movement of point particle in P scheme leads to the renormalization procedure. This procedure occurs by calculating of loops. The problem is that in a loop we have the product of propagators of free particles. Propagators are distributions and consequently their product is not well defined  mathematical procedure. We need to use the renormalization procedure to obtain physically meaningful result.

We think that the renormalization procedure will not appear in NP scheme. But it does not mean that in this case the life will be easier. We will have much more complicated problems: cutting off an infinite equations set for all Green functions, closure problem and so on.

\section{Comparison with turbulence modeling}

In this section we want to compare the procedure of NP quantization with turbulence modeling. In principle, the time-dependent, three-dimensional Navier-Stokes equation contains all of the physics of a given turbulent flow (in this section we follow the textbook~\cite{Wilcox})
  \begin{equation}
    \rho \left(
     \frac{\partial v_i}{\partial t}
    + v_j \cdot \frac{\partial v_i}{\partial x_j}
    \right) = - \frac{\partial p}{\partial x_i} +
    \frac{\partial t_{ij}}{\partial x_j},
  \label{4-10}
  \end{equation}
  where $v_i$ is the flow velocity, $\rho$ is the fluid density, $p$ is the pressure, $t_{ij} = 2 \mu s_{ij}$ is the viscous stress tensor,
$s_{ij} = \frac{1}{2} \left(
    \frac{\partial v_i}{\partial x_j} + \frac{\partial v_j}{\partial x_i}
  \right)$, and $\mu$ is molecular viscosity.

Because turbulence consists of random fluctuations of the various flow properties, a statistical approach for turbulence modeling is used. For a complete statistical description of the hydrodynamic fields of a turbulent flow it is required to have all the multidimensional joint probability distribution for the values of these characteristics in spacetime. But the definition of multivariate distributions is a very complex problem, in addition, these distributions are themselves often inconvenient for applications due to its awkwardness. Therefore, in practice one can form the time average of the continuity and Navier-Stokes equations. The nonlinearity of the Navier-Stokes equation leads to the appearance of momentum fluxes that act as apparent stresses throughout the flow. After that it is necessary to derive equations for these stresses and the resulting equations include additional unknown quantities. This illustrates the issue of closure, i.e., establishing a sufficient number of equations for all of the unknown cumulants (moments).

\subsection{Reynolds Averaging}

One can introduce the instantaneous velocity, $v_i(\vec x, t)$, as the sum of a mean, $V_i(\vec x, t)$, and a fluctuating part, $v^\prime_i(\vec x, t)$, so that
\begin{eqnarray}
  v_i(\vec x, t) &=& V_i(\vec x, t) + v^\prime_i(\vec x, t),
\label{4a-10}\\
  \rho \frac{\partial V_i}{\partial t} +
  \rho V_j \frac{\partial V_i}{\partial x_j}
  &=& - \frac{\partial p}{\partial x_i} +
  \frac{\partial }{\partial x_j} \left(
    2 \mu S_{ji} - \overline{\rho v_j^\prime v_i^\prime}
  \right).
\label{4a-20}
\end{eqnarray}
Equation \eqref{4a-20} is usually referred to as the Reynolds-averaged Navier-Stokes equation. The quantity $\overline{\rho v_j^\prime v_i^\prime}$ is known as the Reynolds-stress tensor and denoted as
\begin{equation}
  \tau_{ij} = - \overline{\rho v_i^\prime v_j^\prime}.
\label{4a-30}
\end{equation}
Immediately we see that we have additionally six unknown quantities $\tau_{ij}$ as a result of Reynolds averaging. It is necessary to note that we have gained no additional equations. For general three-dimensional flows, we have four unknown mean-flow properties: pressure $p$ and the three velocity components $v_i$.

Along with the six Reynolds-stress components, we thus have ten unknown quantities. Our equations are mass conservation and the three components of equation \eqref{4a-20}. This means that our mathematical description of a turbulent flow is not yet closed. To close the system, we must find enough equations to solve for our unknowns.

\subsection{Equation for the Reynolds stress tensor}

To obtain additional equations, we can take moments of the Navier-Stokes equation. That is, we multiply the Navier-Stokes equation by a fluctuating property and average the product. Using this procedure, we can derive, for example, a differential equation for the Reynolds-stress tensor.

One can obtain the following equation for the Reynolds stress tensor
\begin{equation}
\begin{split}
  \frac{\partial \tau_{ij}}{\partial t} +
  V_k \frac{\partial \tau_{ij}}{\partial x_k} =
  - \tau_{ik} \frac{\partial V_j}{\partial x_k} -
  \tau_{jk} \frac{\partial V_i}{\partial x_k} +
  2 \mu \overline{\frac{\partial v^\prime_i}{\partial x_k}
  \frac{\partial v^\prime_j}{\partial x_k}} -
  \overline{
  		p^\prime \left(
  			\frac{\partial v_i^\prime}{\partial x_j} +
  			\frac{\partial v_j^\prime}{\partial x_i}
  		\right)
  } +
  \frac{\partial}{\partial x_k} \left(
    \nu \frac{\partial \tau_{ij}}{\partial x_k} +
    C_{ijk}
    \right),
\label{4b-10}
\end{split}
\end{equation}
where
\begin{equation}
	C_{ijk} =
	\overline{\rho v^\prime_i \rho v^\prime_j \rho v^\prime_k} +
	\delta_{jk} \overline{p^\prime v^\prime_i} +
	\delta_{ik} \overline{p^\prime v^\prime_j}.
\end{equation}
We have obtained six new equations, one for each independent component of the Reynolds-stress tensor $\tau_{ij}$. However, we have also generated 22
new unknowns $\overline{\rho v^\prime_i \rho v^\prime_j \rho v^\prime_k}$, $\overline{\mu \frac{\partial v^\prime_i}{\partial x_k}
\frac{\partial v^\prime_j}{\partial x_k}}$ and
$\overline{v^\prime_i \frac{\partial p^\prime}{\partial x_j}}$,
$\overline{v^\prime_j \frac{\partial p^\prime}{\partial x_i}}$.

It illustrates the closure problem of turbulence. Because of the nonlinearity of the Navier-Stokes equation, as we take higher and higher moments, we generate additional unknowns at each level. The function of turbulence modeling is to devise approximations for the unknown correlations in terms of flow properties that are known so that a sufficient number of equations exists. In making such approximations, we close the system.

There exist a few approaches to close this infinite set of equations. Some of them are: algebraic (zero-equation) models, one-equation models, two-equation models, and second-order closure models.

\textcolor{blue}{\emph{Equations \eqref{4a-20} and \eqref{4b-10} are the first two in the infinite set of equations for all cumulants.
The same holds for
the infinite set of equations for all Green's functions in the case of NP quantization.
The main problem in both cases is the closure problem whose essence is that it is necessary to cut off the infinite system of equations.}}

\subsection{Cutting off in turbulence modeling}

In this subsection we want to take a look over the procedure of cutting off in turbulence modeling.

\subsubsection{Algebraic model}

The simplest of turbulence models is an algebraic model. This model uses the so-called Boussinesq eddy-viscosity approximation for computing the Reynolds stress tensor as the product of an eddy viscosity and the mean strain-rate tensor. Often the eddy viscosity, in turn, is computed in terms of a mixing length. According to the textbook \cite{Wilcox},
\begin{equation}
	\tau_{xy} = \mu_T \frac{d U}{dy},
\label{4c1-10}
\end{equation}
where the eddy viscosity $\mu_T$ is given by
\begin{equation}
	\mu_T = \rho l^2_{mix}\frac{d U}{dy}
\label{4c1-20}
\end{equation}
with Prandtl's empirical mixing length, $l_{mix}$. Let us note that it allows one to cut off an infinite system of equations to one equation.

\subsubsection{Turbulence energy equation models}

In this approach a transport equation is considered. The equation is obtained from \eqref{4b-10} by tracing it
\begin{equation}
	\rho \frac{\partial k}{\partial t} +
	\rho U_i \frac{\partial k}{\partial x_i} =
	\tau_{ij} \frac{\partial U_i}{\partial x_j} -
	\rho \epsilon +
	\frac{\partial}{\partial x_j} \left(
		\mu 	\frac{\partial k}{\partial x_j} -
		\frac{1}{2}
		\overline{\rho v^\prime_i v^\prime_j v^\prime_k} -
		\overline{p^\prime v^\prime_j}
	\right),
\label{4c2-10}
\end{equation}
where $p^\prime$ is the deviation of pressure from the averaged value, and
the specific turbulence kinetic energy $k$ is defined as
\begin{equation}
	2 \rho k = - \tau_{ii}.
\label{4c2-20}
\end{equation}
The dissipation per unit mass $\epsilon$ is defined as
\begin{equation}
	\epsilon = \nu \overline{
		\frac{\partial v^\prime_i}{\partial x_k}	
		\frac{\partial v^\prime_i}{\partial x_k}	
	}.
\label{4c2-30}
\end{equation}
The conventional approach to close of $k$ equation consists in replacing  unknown correlations with closure approximations.

Reynolds stress tensor
\begin{equation}
	\tau_{ij} = 2 \mu_T S_{ij} -
	\frac{2}{3} \rho k \delta_{ij},
\label{4c2-40}
\end{equation}
where $S_{ij}$ is the mean strain-rate tensor.

The standard approximation of a turbulent transport of scalar quantities in a turbulent flow is that of gradient-diffusion. In our case it is
\begin{equation}
	\frac{1}{2} \overline{
		\rho v^\prime_i v^\prime_i v^\prime_j
	} + \overline{p^\prime v^\prime_j} =
	-\frac{\mu_T}{\sigma_k} \frac{\partial k}{\partial x_j},
\label{4c2-50}
\end{equation}
where $\sigma_k$ is a closure coefficient.

To complete equation \eqref{4c2-10}, Prandtl postulated the following form for the dissipation,
\begin{equation}
	\epsilon = \frac{C_D}{l_{mix}} k^{3/2},
\label{4c2-60}
\end{equation}
where $C_D$ is a closure coefficient and the turbulence scale length remains the only indefinite parameter in this model. Thus, combining equations \eqref{4c2-10} and \eqref{4c2-60}, the one-equation model appears as
\begin{equation}
	\rho \frac{\partial k}{\partial t} +
	\rho U_i \frac{\partial k}{\partial x_i} =
	\tau_{ij} \frac{\partial U_i}{\partial x_j} -
	\frac{C_D}{l_{mix}} k^{3/2} +
	\frac{\partial}{\partial x_j} \left[
		\left(
      \mu + \frac{\mu_T}{\sigma_k}
    \right) \frac{\partial k}{\partial x_j}
	\right].
\label{4c2-70}
\end{equation}

\subsubsection{Two-equation models}

The starting point for two-equation models is the Boussinesq approximation \eqref{4c2-40} and the turbulence kinetic energy equation \eqref{4c2-70}. Kolmogorov pointed out~\cite{kolmogorov}
that a second transport equation is needed to compute the specific dissipation rate, $\omega$. There are many two-equation models. For example, we consider the $k-\omega$ model.

Kolmogorov postulated~\cite{kolmogorov} the following equation for $\omega$,
\begin{equation}
	\rho \frac{\partial \omega}{\partial t} +
	\rho U_i \frac{\partial \omega}{\partial x_i} =
	- \beta \rho \omega^2 +
  \frac{\partial}{\partial x_j} \left(
		\sigma \mu_T \frac{\partial \omega}{\partial x_j}
	\right),
\label{4d2-10}
\end{equation}
where $\beta$ and $\sigma$ are two new closure coefficients. Finally, we have two equations for a turbulent fluid:
\begin{eqnarray}
  \rho \frac{\partial k}{\partial t} +
	\rho U_i \frac{\partial k}{\partial x_i} &=&
	\tau_{ij} \frac{\partial U_i}{\partial x_j} -
	\beta^* \rho k \omega +
	\frac{\partial}{\partial x_j} \left[
		\left(
      \mu + \sigma^* \mu_T
    \right) \frac{\partial k}{\partial x_j}
	\right],
\label{4d2-20}\\
  \rho \frac{\partial \omega}{\partial t} +
	\rho U_i \frac{\partial \omega}{\partial x_i} &=&
  \alpha \frac{\omega}{k} \tau_{ij} \frac{\partial U_i}{\partial x_j}
	- \beta \rho \omega^2 +
  \frac{\partial}{\partial x_j} \left[
		\left(
      \mu + \sigma \mu_T
    \right) \frac{\partial \omega}{\partial x_j}
	\right],
\label{4d2-30}\\
  \mu_T &=& \frac{\rho k}{\omega},
\label{4d2-40}\\
  \epsilon &=& \beta^* \omega k , \;
  l_{mix} = \frac{k^{1/2}}{\omega},
\label{4d2-50}
\end{eqnarray}
where \eqref{4d2-20} is the turbulence kinetic energy equation; \eqref{4d2-30} is the specific dissipation rate equation; \eqref{4d2-40} is the eddy viscosity; $\alpha, \beta, \beta^*, \sigma, \sigma^*$ are closure coefficients; \eqref{4d2-50} are auxiliary relations. It is necessary to note that there are many other two-equation models (see the textbook~\cite{Wilcox} for details).

It is important for us that in such a case \textcolor{blue}{\emph{we cut off an infinite set of equations for cumulants to two equations with some closure coefficients.}}

\subsection{Closure coefficients problem}

As we have seen from the above discussion, the cutoff of infinite sets of equations in turbulence modeling is performed
 by using physical assumptions about higher order cumulants and by introducing closure coefficients.
It is written in Ref.~\cite{Wilcox}:
\textcolor{blue}{
``All of the two-equation models have closure coefficients that have been introduced in replacing unknown double and triple correlations with algebraic expressions involving known turbulence and mean-flow properties. The Wilcox $k-\omega$ model, for example, has five viz.,
$\alpha, \beta, \beta^*, \sigma, \sigma^*$. If our theory were exact, we could set the values of these coefficients from first principles much as we use the kinetic theory of gases to determine the viscosity coefficients is Stokes' approximation for laminar flows. Unfortunately, the theory is not exact, but rather a model developed mainly on the strength of dimensional analysis. Consequently, the best we can do is to set the values of the closure coefficients to assure agreement with observed properties of turbulence.''}

This statement can be directly transferred to a NP quantization procedure by replacing the word ``turbulence'' by  ``quantum''.

\subsection{Reynolds number and coupling constant}

In this section we want to discuss the connection between turbulence and quantum chromodynamics (for details, see Ref.~\cite{Dzhunushaliev:2009turb-qcd}).

The character of a fluid flow depends on the Reynolds number $\mathrm{Re}$
\begin{equation}
	\mathrm{Re} = \frac{\rho_m v l}{\mu},
\label{comp-e-10}
\end{equation}
where $\rho_m$ is the fluid density, $v$ is the fluid velocity, $l$ is the characteristic length for a given flow, and $\mu$ is the fluid viscosity. If $\mathrm{Re} < \mathrm{Re}_{cr}$ then the flow is the laminar one, if $\mathrm{Re} > \mathrm{Re}_{cr}$ -- the turbulent one. The creation of turbulence at
$\mathrm{Re} \approx \mathrm{Re}_{cr}$ is an open question in hydrodynamics.

In quantum field theory there are the perturbative regime when a dimensionless coupling constant
\begin{equation}
	\alpha^2 = \frac{1/\tilde g^2}{\hbar c}
\label{comp-e-20}
\end{equation}
is small enough, $\alpha^2 < 1$, and the nonperturbative regime when $\alpha^2 \geq 1$ (here $\tilde g$ is a dimension coupling constant, $\hbar$
is the Planck constant, and $c$ is the speed of light). In quantum field theory a nonperturbative quantization for $\alpha^2 \geq 1$ is an open question.

Possible connection between hydrodynamics and quantum field theory based on a comparison between the Reynolds number (in hydrodynamics) and
a dimensionless coupling constant (in quantum field theory) is proposed in Ref.~\cite{Dzhunushaliev:2009turb-qcd}. Let us rewrite Eq. \eqref{comp-e-10} in the following form:
\begin{equation}
	\mathrm{Re} = \frac{\rho v^2 l^4}{\mu l^3 v}
\label{comp-e-30}
\end{equation}
with
$\left[ \rho v^2 l^4 \right] = \left[ 1/\tilde g^2 \right] = \text{g} \cdot \text{cm}^3 / \text{s}^2$,
$\left[ \hbar \right] = \left[ \mu l^3 \right] = \text{g} \cdot \text{cm}^3 / \text{s}$. Now assume that there are the following relations:
\begin{eqnarray}
	1/\tilde g^2 & \leftrightarrow & \rho v^2 l^4 ,
\label{comp-e-40}\\
	\hbar & \leftrightarrow & \mu l^3 .
\label{comp-e-50}
\end{eqnarray}
This leads to the following correspondence:
\begin{itemize}
	\item $\hbar = 0 \leftrightarrow \mu = 0$: a classical theory corresponds to an ideal fluid.
	\item $\hbar \neq 0 \text{ and } \alpha^2 < 1 \leftrightarrow \mu \neq 0 \text{ and } \mathrm{Re} < \mathrm{Re}_{cr}$: a laminar fluid corresponds to a perturbative regime of quantum field theory.
	\item $\hbar \neq 0 \text{ and } \alpha^2 \geq 1 \leftrightarrow \mu \neq 0 \text{ and } \mathrm{Re} > \mathrm{Re}_{cr}$: a turbulent fluid corresponds to a nonperturbative regime of quantum field theory.
\end{itemize}

\section{Perspectives and applications}

In this section we want to give examples how a NP quantization can be applied  in certain cases.

\subsection{Gravity}

Here we consider a few cases of NP quantization as applied to gravity.

\subsubsection{Modified gravity from the quantum part of the metric. Part A}
\label{parta}

We recall that a NP quantization for gravity is the operator Einstein equation
\begin{equation}
	\hat R_{\mu \nu} - \frac{1}{2} \hat g_{\mu \nu} \hat R =
	\varkappa \hat T_{\mu \nu} .
\label{app-gr-10}
\end{equation}
In this subsection we want to consider a physical system where the quantum metric $\hat{\mathfrak g}_{\mu \nu}$ can be decomposed into two parts: the classical metric $g_{\mu \nu}$ and the quantum metric
$\widehat g_{\mu \nu}$ (here we follow Ref.~\cite{Dzhunushaliev:2013nea}),
\begin{equation}
  \hat{\mathfrak g}_{\mu \nu} = g_{\mu \nu} + \widehat g_{\mu \nu},
\label{app-gr-20}
\end{equation}
together with the assumption that
\begin{equation}
  \left\langle \widehat g_{\mu \nu} \right\rangle \neq 0.
\label{app-gr-30}
\end{equation}
In order to avoid an infinite set of equations for all Green's functions,
 we want to write the averaged Lagrangian using the decomposition \eqref{app-gr-20} with small $\widehat g_{\mu \nu}$,
\begin{equation}
  \left\langle \widehat g_{\mu \nu} \right\rangle \ll
  g_{\mu \nu} .
\label{app-gr-40}
\end{equation}
We start from the Einstein-Hilbert Lagrangian
\begin{equation}
\mathcal L_{\mathcal G} = - \frac{c^2}{2 \varkappa} \sqrt{-\mathcal G} R,
\label{app-gr-50}
\end{equation}
where $\varkappa=8\pi G/c^2$. We use some physically reasonable assumptions about the expectation value  $\left\langle \widehat{g}^{\mu \nu} \right\rangle$,
\begin{equation}
	\left\langle \widehat{g}_{\mu \nu} \right\rangle =
  K_{\mu \nu} \approx
   F(R, R_{\mu \nu} R^{\mu \nu}, \cdots) \times
   \begin{cases}
    \text{either} & g_{\mu \nu} \\
    \text{or}     & R_{\mu \nu}
  \end{cases},
\label{app-gr-60}
\end{equation}
where $F(R, R_{\mu \nu} R^{\mu \nu}, \cdots)$ is the proportionality coefficient (closure coefficient in the language of turbulence modeling).
With such assumptions about the expectation value of the quantum part $\widehat g_{\mu \nu}$ of the metric
$\mathcal G_{\mu \nu}$ the Lagrangian \eqref{app-gr-50} takes the form
\begin{equation}
  \left\langle 	
		\mathcal L_{\mathcal G}(g + \widehat g)
	\right\rangle \approx - \frac{c^2}{2 \varkappa} \sqrt{-g}
  \left(
    R + G_{\mu \nu} K^{\mu \nu}
  \right).
\label{app-gr-70}
\end{equation}
\textcolor{blue}{
\emph{Hence we see that the quantum corrections coming from a nonzero expectation value of the nonperturbatively quantized metric give rise to modified gravity theories. }}

In such a physical system quantum corrections for matter are also appear.
For example, the scalar field Lagrangian changes as follows:
\begin{equation}
	\left\langle \mathcal L_m^{\mathcal G}(g + \widehat{g}) \right\rangle =
  \sqrt{-g}\left[\frac{1}{2}\nabla^\mu \phi \nabla_\mu \phi - V(\phi)
 + \frac{1}{2}T_{\mu \nu} K^{\mu \nu}\right],
\label{app-gr-80}
\end{equation}
where $T_{\mu \nu}$ is the energy-momentum tensor. \textcolor{blue}{\emph{Thus, we have obtained a scalar field nonminimally coupled to gravity.}}

\subsubsection{Modified gravity from the quantum part of the metric. Part B}

In the previous subsection \ref{parta} we have considered an approximate decomposition of the quantum metric \eqref{app-gr-20} where, in the first approximation, the metric operator is decomposed into the classical part and the quantum part with zero expectation value. One can consider a case where the quantum expectation value is nonzero and can be decomposed
into two parts (here we follow Ref. \cite{Dzhunushaliev:2015}):
\begin{equation}
	\hat g_{\mu\nu} \approx g_{\mu \nu} + \widehat{\delta g}_{\mu \nu} +
	\widehat{\delta^2 g}_{\mu \nu},
\label{partb-10}
\end{equation}
where
$
\left\langle Q \left|
	\widehat{\delta g}_{\mu \nu}
\right | Q \right\rangle = 0
$;
$
\left\langle Q \left|
	\widehat{\delta^2 g}_{\mu \nu}
\right | Q \right\rangle \neq 0
$;
$\widehat{\delta g}_{\mu \nu}$ and $\widehat{\delta^2 g}_{\mu \nu}$
are the first- and second-order deviations of the operator $\hat g_{\mu\nu}$.

We use the Einstein-Hilbert Lagrangian \eqref{app-gr-50}
\begin{equation}
	\mathcal L(g + \widehat{\delta g}+\widehat{\delta^2 g}) \approx \mathcal L(g) +
	\frac{\delta \mathcal L}{\delta g^{\mu \nu}} \widehat{\delta g^{\mu \nu}} +
	\frac{\delta^2 \mathcal L}{\delta g^{\mu \nu} \delta g^{\rho \sigma}}
	\widehat{\delta g^{\mu \nu}} \widehat{\delta g^{\rho \sigma}} +
	\frac{\delta^2 \mathcal L}{\delta^2 g^{\mu \nu}}
	\widehat{\delta^2 g^{\mu \nu}} .
\label{partb-20}
\end{equation}
We need some assumption about a 2-point Green's function: we suppose that it can be decomposed as the product
\begin{equation}
	G_{2; \mu \nu, \rho \sigma} \left( x_1, x_2 \right) =
	\left\langle Q \left|
		\widehat{\delta g_{\mu \nu}}(x_1) \cdot
    \widehat{\delta g_{\rho \sigma}}(x_2)
	\right | Q \right\rangle \approx
	P_{\mu \nu}(x_1) P_{\rho \sigma}(x_2)
\label{partb-30}
\end{equation}
with
\begin{equation}
  \left\langle \widehat{\delta^2 g_{\mu \nu}} \right\rangle
	= K_{\mu \nu}.
\label{parb-40}
\end{equation}
The simplest assumption about $P_{\mu \nu}$ is
\begin{equation}
	P_{\mu \nu} = F(R, R_{\mu \nu} R^{\mu \nu}, \cdots) g_{\mu \nu}.
\label{partb-50}
\end{equation}
That gives us the following modified gravity
\begin{equation}
	\left\langle \mathcal L(g + \widehat{\delta g}+\widehat{\delta^2 g}) \right\rangle \approx
	- \frac{c^2}{2 \varkappa} \sqrt{-g} \left[ R -
	2 R F(R, \cdots) + 3 F(R, \cdots) \nabla^\mu \nabla_\mu F(R, \cdots) +
  G_{\mu \nu} K^{\mu \nu}
	\right].
\label{partb-60}
\end{equation}
Similar calculations for a scalar field Lagrangian  yield
\begin{equation}
	\left\langle
    \mathcal L_m + \widehat{\delta^2 \mathcal L_m}
  \right\rangle =
	\sqrt{-g}\left\{
		\frac{1}{2}
		 \nabla^\mu \phi \nabla_\mu \phi -
		\left[
			1 + 2 F\left( R, \cdots \right)
		\right] V(\phi) +
		K^{\mu \nu} T_{\mu \nu}
	\right\} .
\label{partb-70}
\end{equation}
Thus, we see that the nonminimal coupling between the scalar field and gravity appears.

\subsubsection{Cosmological constant and Euclidean space from nonperturbative quantum torsion}

In this subsection we want to show that in a gravitating physical system a situation may occur where a fluctuating torsion gives rise to
the appearance of
a cosmological constant and Euclidezation of spacetime (here we follow Ref.~\cite{Dzhunushaliev:2012nf}). We consider the case with the classical metric, $g_{\mu \nu}$,
and the quantum affine connection,
$
\hat \Gamma^\rho_{\phantom{\rho} \mu \nu} =
	\hat G^\rho_{\phantom{\rho} \mu \nu} +
	\hat K^\rho_{\phantom{\rho} \mu \nu}
$,
where the Christoffel symbols (denoted by  $G$) are in a classical mode and the contorsion tensor $\hat K$ is in a quantum mode.

Let us assume that the expectation value of the contortion tensor is zero,
$
\left\langle
		\hat K^\rho_{\phantom{\rho} \mu \nu}
	\right\rangle = 0,
$ and consequently
\begin{equation}
	\left\langle
		\hat \Gamma^\rho_{\phantom{\rho} \mu \nu}
	\right\rangle =
	G^\rho_{\phantom{\rho} \mu \nu},
\label{app-ts-10}
\end{equation}
but the standard deviation of the contortion tensor is not zero:
\begin{equation}
	\left\langle
		\left( \hat K^\rho_{\phantom{\rho} \mu \nu} \right)^2
	\right\rangle \neq 0.
\label{app-ts-20}
\end{equation}
For simplicity, we also assume that the contortion tensor is absolutely antisymmetric. In this case
\begin{equation}
	\hat K_{\rho \mu \nu} = \hat Q_{\rho \mu \nu} = \hat Q_{[\rho \mu \nu]},
\label{app-ts-30}
\end{equation}
where $\hat Q_{\rho \mu \nu}$ is the torsion tensor operator.

Again our strategy is not to solve an infinite set of
equations  for all Green's functions. Instead of this, we will average the
Einstein equations (that is analogous to averaging of the Navier-Stokes equation in turbulence modeling).
Both fields describe approximately the dispersion of quantum fluctuation of the torsion.

\textbf{Scalar field approximation. } For the scalar field approximation
\begin{equation}
	\left\langle
		\hat Q_{\rho_1 \mu_1 \nu_1}(x) \hat Q_{\rho_2 \mu_2 \nu_2}(x)
	\right\rangle \approx \varsigma
	\varepsilon_{\rho_1 \mu_1 \nu_1 \alpha}(x)
	\varepsilon_{\rho_2 \mu_2 \nu_2 \beta}(x)
	g^{\alpha \beta}(x)
	\left| \phi(x) \right|^2,
\label{app-ts-40}
\end{equation}
where $\varepsilon_{\rho \mu \nu \sigma}$ is the absolutely antisymmetric  Levi-Civita tensor and $\phi(x)$ is a scalar field.

After some calculations the averaged vacuum Einstein equations with nonperturbative quantum gravitational admixtures are
\begin{equation}
	R_{\mu \nu} - \frac{1}{2} g_{\mu \nu} R -
	6 \varsigma g_{\mu \nu} \left| \phi \right|^2 = 0.
\label{app-ts-50}
\end{equation}
In order to obtain an equation for the scalar field, we use the Bianchi identities for the vacuum Einstein equations:
\begin{equation}
	\left\langle
	\left(
		\hat{R}^\mu_\nu - \frac{1}{2} \delta^\mu_\nu \hat R
	\right)_{; \mu}
	\right\rangle = 0.
\label{app-ts-60}
\end{equation}
The desired equation for $\phi$ is
\begin{equation}
	\left| \phi \right|_{;\mu} = \left| \phi \right|_{, \mu} = 0 .
\label{app-ts-70}
\end{equation}
Equation \eqref{app-ts-70} gives us the following solution:
\begin{equation}
	\phi = \mathrm{const},
\label{app-ts-80}
\end{equation}
and we can identify the scalar field describing the nonperturbative quantum effects with a cosmological constant in the following way:
\begin{equation}
	\Lambda = - 6 \varsigma \left| \phi \right|^2.
\label{app-ts-90}
\end{equation}

\textbf{Vector field approximation. } For the vector field approximation we assume that
\begin{equation}
	\left\langle
		\hat Q_{\rho_1 \mu_1 \nu_1}(x_1) \hat Q_{\rho_2 \mu_2 \nu_2}(x_2)
	\right\rangle \approx
  \varepsilon_{\rho_1 \mu_1 \nu_1 \alpha}(x_1)
	\varepsilon_{\rho_2 \mu_2 \nu_2 \beta}(x_2)
	A^\alpha (x_1) A^\beta (x_2).
\label{app-ts-100}
\end{equation}
Here the vector field $A^\alpha$ describes the dispersion of the torsion. The vacuum Einstein equations are
\begin{equation}
	\tilde{R}_{\mu \nu} -
		\frac{1}{2} g_{\mu \nu} \tilde{R} - \varsigma \left(
		g_{\mu \nu} A_\alpha A^\alpha + 2 A_\mu A_\nu
	\right) = 0.
\label{app-ts-110}
\end{equation}
From the requirement that the Bianchi identities \eqref{app-ts-60} are satisfied, we obtain the following equation for the vector field $A_\mu $:
\begin{equation}
	\left(
		\delta^\mu_\nu A^\alpha A_\alpha + 2 A^\mu A_\nu
	\right)_{; \mu} = 0.
\label{app-ts-120}
\end{equation}
Let us consider the spherically symmetric metric
\begin{equation}
	ds^2 = b^2(r) \Delta(r) dt^2 -
	\frac{dr^2}{\Delta(r)} -
	r^2 \left(
		d \theta^2 + \sin^2 \theta d \varphi^2
	\right)
\label{app-ts-130}
\end{equation}
with the vector $A_\mu$
\begin{equation}
	A_\mu = \left( \phi(r), 0, 0, 0 \right).
\label{app-ts-140}
\end{equation}
The corresponding Einstein equations and equation for $A_\mu$  \eqref{app-ts-120} give us the following solution
\begin{equation}
	ds^2 = - dt^2 - l_0^2 \left[
	d\chi^2 + \sinh^2 \chi \left(
		d \theta^2 + \sin^2 \theta d \varphi^2
	\right)
	\right].
\label{app-ts-150}
\end{equation}
Here we introduced new coordinate $\chi = \mathrm{arcsinh} \frac{r}{l_0}$. It is interesting to note that $g_{tt} = - 1$ is negative and $t$ becomes an imaginary time.

Finally, an interesting result is that the vacuum Einstein equations give us the regular (but Euclidean) solution.
The reason for this is that nonperturbative quantum gravitational effects are taken into account.

\subsection{Quantum chromodynamics}

In this section we want to consider a few examples of applying NP quantization ideas for quantum chromodynamics.

\subsubsection{Scalar toy model of glueball}
\label{glueball}

A glueball is a hypothetical composite particle that consists solely of non-Abelian SU(3) gauge field, without valence quarks. The existence of a glueball is a consequence of the self-interaction of gluons within quantum chromodynamics (QCD). Nonlinear self-interaction of gluons in QCD leads to possibility of the existence of a color-neutral state made of gluons only, which was called glueball. Glueball is also thought of as a bound state of gluons, and it's properties cannot be described within a perturbative approach to QCD. Glueball remains an obscure object over thirty years after QCD was used to predict such a state. It is well known that the gluon condensate, from which gluball is thought to be made of,  can only be determined in a nonperturbative formulation of QCD. So far various attempts have been made to determine gluon condensate from first principles \cite{Banks}, \cite{Shifman:1978bx}. We refer the reader to Ref.~\cite{Mathieu:2008me} for more details.

Again, instead of solving infinite set of equations for all Green's functions, we want to write an effective Lagrangian. This Lagrangian is obtained from the Lagrangian of SU(3) non-Abelian gauge theory. In order to do this, we first separate SU(3) color degrees of freedom into two parts: subgroup SU(2) $\subset$ SU(3) and coset SU(3)/SU(2). Then we average the SU(3) Lagrangian using some assumptions and approximation. Our approximation is based on the main assumption that the 2 and 4-points Green's functions are described in terms of some scalar fields $\phi$ and $\chi$ due to the following relations:
\begin{align}
	\left( G_2 \right)^{ab}_{\mu \nu}(x_1, x_2) & =
	\left\langle
		A^a_\mu (x_1) A^b_\nu (x_2)
	\right\rangle
	& \approx & \;
	C^{ab}_{\mu \nu} \phi(x_1) \phi^*(x_2) +
	m^{ab}_{\mu \nu},
\label{qcd-gl-10}\\
	\left( G_4 \right)^{abcd}_{\mu \nu \rho \sigma}(x_1, x_2, x_3, x_4) &=
	\left\langle
		A^a_\mu (x_1) A^b_\nu (x_2) A^c_\rho (x_3) A^d_\sigma (x_4)
	\right\rangle
	&\approx & \;
	\left\langle
		A^a_\mu (x_1) A^b_\nu (x_2)
	\right\rangle
	\left\langle
		A^c_\rho (x_3) A^d_\sigma (x_4)
	\right\rangle
\label{qcd-gl-20}\\
	\left( G_2 \right)^{mn}_{\mu \nu}(x_1, x_2) &=
	\left\langle
		A^m_\mu (x_1) A^n_\nu (x_2)
	\right\rangle
	&\approx & \;
	C^{mn}_{\mu \nu} \chi(x_1) \chi^*(x_2) +
	m^{mn}_{\mu \nu},
\label{qcd-gl-30}\\
	\left( G_4 \right)^{mnpq}_{\mu \nu \rho \sigma}(x_1, x_2, x_3, x_4) &=
	\left\langle
		A^m_\mu (x_1) A^n_\nu (x_2) A^p_\rho (x_3) A^q_\sigma (x_4)
	\right\rangle
	&\approx & \;
	\left\langle
		A^m_\mu (x_1) A^n_\nu (x_2)
	\right\rangle
	\left\langle
		A^p_\rho (x_3) A^q_\sigma (x_4)
	\right\rangle,
\label{qcd-gl-40}
\end{align}
where $a,b,c,d = 1,2,3$ are SU(2) indices, $m,n,p,q = 4,5,6,7,8$ are coset indices, and $C^{ab}_{\mu\nu}$ and $m^{ab}_{\mu\nu}$ are closure constants. We see that similarly to turbulence modeling we have to introduce some closure constants.

An effective Lagrangian then becomes
\begin{equation}
	\mathcal L_{\rm{eff}} = \left\langle \mathcal L_{SU(3)} \right\rangle =
	\frac{1}{2} \left| \nabla_\mu \phi \right|^2 -
	\frac{\lambda_1}{4} \left(
		\left| \phi \right|^2 - \phi_\infty^2
	\right)^2 +
	\frac{1}{2} \left| \nabla_\mu \chi \right|^2 -
	\frac{\lambda_2}{4} \left(
		\left| \chi \right|^2 - \chi_\infty^2
	\right)^2 + \frac{\lambda_2}{4} \chi^4_\infty -
	\frac{1}{2} \left| \phi \right|^2 \left| \chi \right|^2,
\label{qcd-gl-50}
\end{equation}
where $\lambda_{1,2}$, $\phi_\infty$, and $\chi_\infty$ are some parameters; the signature of the spacetime metrics is $(+,-,-,-)$. The effective Lagrangian \eqref{qcd-gl-50} is an approximation to the nonperturbatively quantized SU(3) gauge theory.

Entities entering the Lagrangian \eqref{qcd-gl-50} have the following meanings and origins:
\begin{itemize}
	\item the scalar fields $\phi$ and $\chi$ describe nonperturbatively quantized SU(2) and coset SU(3)/SU(2) degrees of freedom, correspondingly;
	\item the terms $\left| \nabla_\mu \phi \right|^2$ and
	$\left| \nabla_\mu \chi \right|^2$ are the result of the nonperturbative quantum averaging of
	$( \nabla_\mu A^B_\nu )^2$ in the initial SU(3) Lagrangian;
	\item the terms $\phi^4$ and $\chi^4$ are the result of the nonperturbative quantum averaging of $f^{ABC} f^{AMN} A^B_\mu A^C_\nu A^{M \mu} A^{N \nu}$;
	\item the term $\phi^2 \chi^2$  is the result of the nonperturbative quantum averaging of $f^{Aab} f^{Amn} A^a_\mu A^b_\nu A^{m \mu} A^{n \nu}$;
  \item closure coefficients $\phi^4_\infty$ and $\chi^4_\infty$ appear;
	\item the terms $\phi^2 \phi^2_\infty$,	$\chi^2 \chi^2_\infty$ arise due to the closure coefficients $\phi_\infty$ and $\chi_\infty$.
\end{itemize}
Using the Lagrangian \eqref{qcd-gl-50}, we derive the associated field equations describing a gluon condensate in the following form:
\begin{eqnarray}
  \partial_\mu \partial^\mu \phi &=&
  - \phi \left[ \left|\chi\right|^2 + \lambda_1
  \left(
    \left|\phi\right|^2 - \phi_\infty
  \right) \right],
\label{qcd-gl-60}\\
  \partial_\mu \partial^\mu \chi &=&
  - \chi \left[ \left|\phi\right|^2 + \lambda_2
  \left(
    \left|\chi\right|^2 - \chi_\infty
  \right) \right].
\label{qcd-gl-70}
\end{eqnarray}
In Ref. \cite{Dzhunushaliev:2003sq} the spherically symmetric solution to equations
 \eqref{qcd-gl-60} and \eqref{qcd-gl-70} is considered. This ball serves us as a nonperturbative scalar model of a glueball.
It is shown in Ref.~\cite{Dzhunushaliev:2003sq}  that such solutions with finite energy do exist.

\subsubsection{Flux tube}

In section we want to show that, using Green's decomposition analogously to the section \ref{glueball}, one can obtain a flux tube stretched between quark and antiquark located at the $\pm$ infinities.

\begin{figure}[h]
  \begin{center}
  \fbox{
  \includegraphics[width=.5\linewidth]{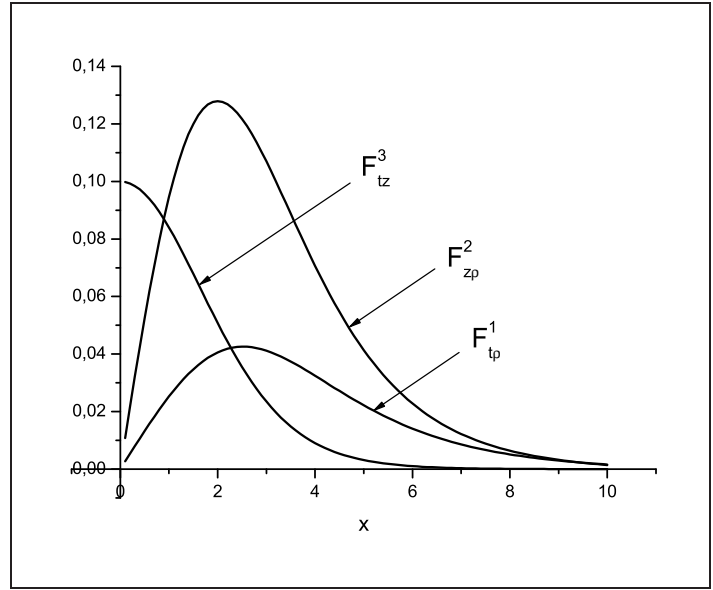}}
  \caption{The profiles of the color fields
  $F^1_{t\rho}(x) = -f', F^4_{z\rho}(x) = -v', F^7_{tz}(x) = f(x) v(x)$.}
  \label{fig2}
  \end{center}
\end{figure}

It is shown in Ref. \cite{Dzhunushaliev:2003sq} that using the following assumptions:
\begin{enumerate}
  \item The SU(3) gauge potential $A^B_\mu \in SU(3), B=1,2, \cdots , 8$ can be separated into two parts:
\begin{itemize}
	\item the first one is the gauge components $A^b_\mu \in SU(2) \subset SU(3)$, which is in a classical state ($b=1,2,3$);
	\item the second one is $A^m_\mu \in SU(3)/SU(2)$, and it is in a quantum state ($m=4,5,6,7,8$).
\end{itemize}
  \item The 2-point Green's function of color quantum field can be approximately presented as the product of scalar fields with some closure coefficients having color and Lorentzian indices:
  \begin{equation}
	 \left( G_2 \right)^{mn}_{\mu \nu}(x_1, x_2) =
	 \left\langle
		A^m_\mu (x_1) A^n_\nu (x_2)
	 \right\rangle \approx C^{mn}_{\mu \nu} \phi(x_1) \phi^*(x_2) +
	\tilde m^{mn}_{\mu \nu}.
  \label{qcd-gl-80}
  \end{equation}
  \item The 4-point Green's function can be decomposed as the product of two 2-point Green's functions:
  \begin{equation}
	 G^{mnpq}_{\mu \nu \rho \sigma}(x_1, x_2, x_3, x_4) =
	 \left\langle
		A^m_\mu (x_1) A^n_\nu (x_2) A^p_\rho (x_3) A^q_\sigma (x_4)
	 \right\rangle \approx
	\left( G_2 \right)^{mn}_{\mu \nu}(x_1, x_2)
    \left( G_2 \right)^{pq}_{\rho \sigma}(x_1, x_2),
  \label{qcd-gl-90}
  \end{equation}
\end{enumerate}
one can obtain an effective Lagrangian
\begin{equation}
	\mathcal L_{eff} = - \frac{1}{4 g^2}
	\left\langle
		\hat {\mathcal F}^{B\mu\nu} \hat {\mathcal F}^{B}_{\mu\nu}
	\right\rangle =
	- \frac{1}{4 g^2} F^{a}_{\mu\nu} F^{a \mu\nu} +
	\frac{1}{2} \left| \partial_\mu \phi \right|^2 -
	\frac{\lambda}{4} \left(
		\left| \phi \right|^2 - \phi_\infty^2
	\right)^2 +
	\frac{1}{2} A^a_\mu A^{a \mu}  \left| \phi \right|^2 -
	\frac{1}{2} m^2_a A^a_\nu A^{a \mu},
\label{qcd-gl-100}
\end{equation}
where $\phi$ is a complex scalar field describing quantum fluctuations of  $A^m_\mu \in SU(3)/SU(2)$ gauge field components; $\phi_\infty, \lambda$, and $m_a$ are closure constants. The field equations for the Lagrangian \eqref{qcd-gl-100} are
\begin{eqnarray}
  \frac{1}{4g^2}D_\nu F^{a\mu\nu} &=& \left(
  	\left| \phi \right|^2 - m^2_a
  \right) A^{a \mu}, \text{ no summation over } a ,
\label{qcd-gl-110}\\
  \Box \phi &=& -\lambda \phi
  \left( \left| \phi \right|^2 - \phi_\infty^2
  \right) + A^a_\mu A^{a \mu} \phi .
\label{qcd-gl-120}
\end{eqnarray}
The flux solution can be sought in the following form:
\begin{equation}
    A^1_t(\rho) = f(\rho) ; \quad A^2_z(\rho) = v(\rho) ;
    \quad \phi(\rho) = \phi(\rho).
\label{qcd-gl-130}
\end{equation}
Here $z, \rho , \varphi$ refer to the cylindrical coordinate system. Substitution of \eqref{qcd-gl-130}
in equations \eqref{qcd-gl-110} and \eqref{qcd-gl-120} gives
\begin{eqnarray}
    f'' + \frac{f'}{x} &=& f \left( \phi^2 + v^2 - m^2_1 \right),
\label{qcd-gl-140}\\
    v'' + \frac{v'}{x} &=& v \left( \phi^2 - f^2 - m^2_2 \right),
\label{qcd-gl-150}\\
    \phi'' + \frac{\phi'}{x} &=& \phi \left[ - f^2 + v^2
    + \lambda \left( \phi^2 - \mu^2 \right)\right].
\label{qcd-gl-160}
\end{eqnarray}
The color electric and magnetic fields are
\begin{equation}
  F^1_{t \rho} = -f', \quad F^2_{z \rho} = - v', \quad
  F^3_{tz} = fv,
\label{qcd-gl-170}
\end{equation}
and they are presented in Fig.~\ref{fig2}.
Let us note that  the color longitudinal electric field $F^3_{tz}$ is not a gradient of a gauge potential but is a consequence of a nonlinear term in the definition of the field tensor $F^a_{\mu \nu}$.

\subsubsection{The gluon condensate distribution in the flux tube}

\begin{figure}[h]
\begin{minipage}[t]{.45\linewidth}
  \begin{center}
  \fbox{
  \includegraphics[width=.95\linewidth,height=0.97\linewidth]{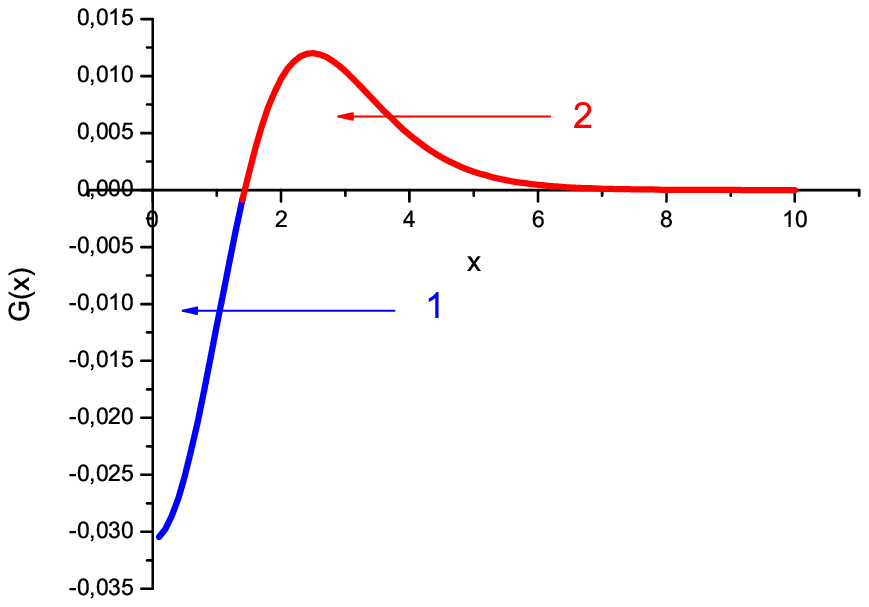}}
  \caption{The SU(3) gluon condensate $G(x)$. 1 is the area where the chromoelectric fields are predominant. 2 is the area where the chromomagnetic fields are predominant.}
  \label{fig3}
  \end{center}
\end{minipage}\hfill
\begin{minipage}[t]{.45\linewidth}
  \begin{center}
  \fbox{
  \includegraphics[width=.85\linewidth]{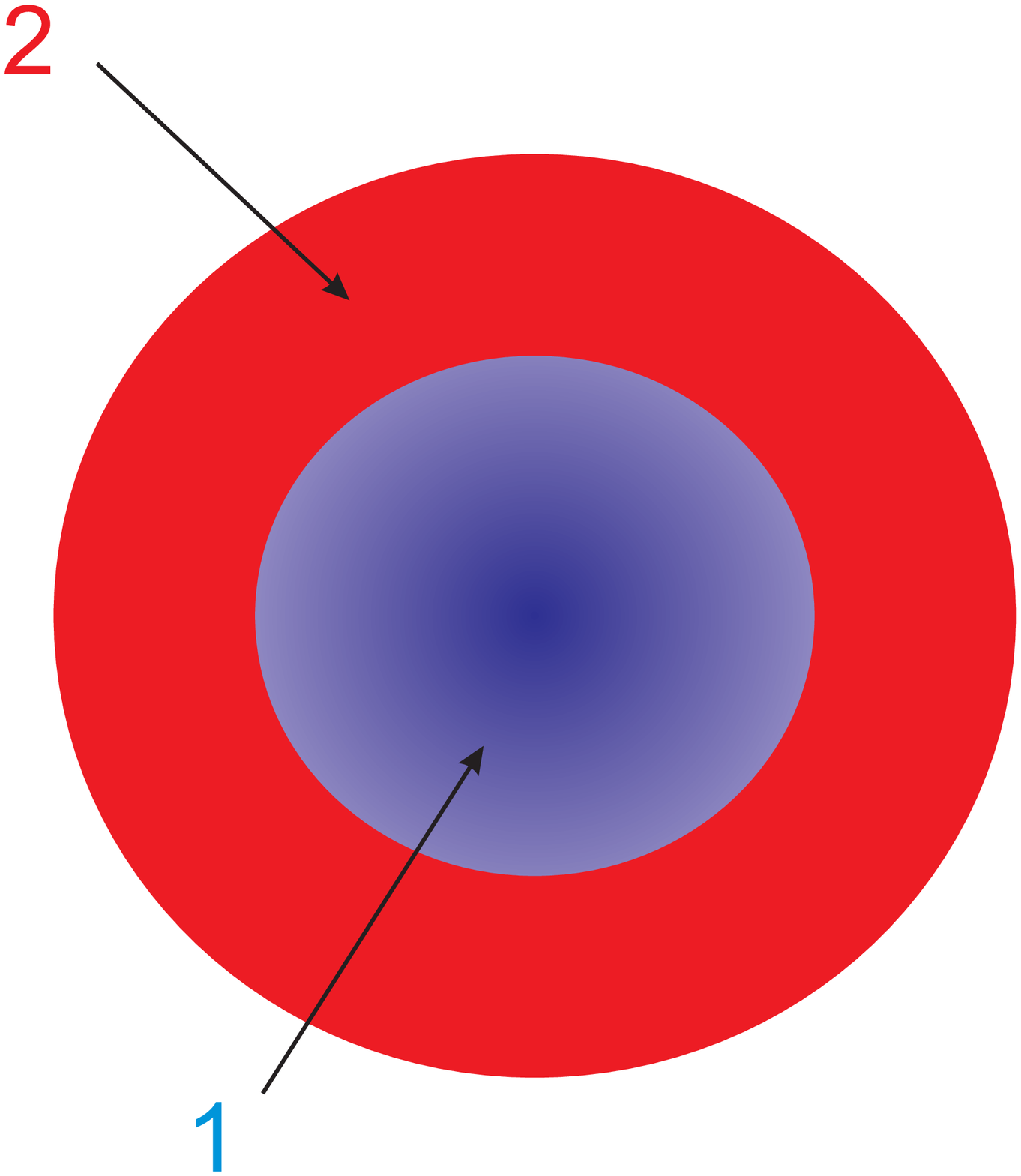}}
  \caption{Sketchy description of the distributions of chromoelectric and chromomagnetic fields: (1) \textcolor{blue}{blue} is the region where the chromoelectric field is predominant and (2) \textcolor{red}{red} is the region where the chromomagnetic field is predominant.}
  \label{fig4}
  \end{center}
\end{minipage}\hfill
\end{figure}

The distribution of the gluon condensate
$\left\langle \mathcal F^A_{\mu\nu} \mathcal F^{A\mu\nu} \right\rangle$ in the flux tube can be found from the effective Lagrangian \eqref{qcd-gl-100}. The gluon condensate is
(here we follow Ref.~\cite{Dzhunushaliev:2010qs})
\begin{equation}
	G = - \mathcal L_{eff} =
	\left\langle \mathcal H^A_\mu \mathcal H^{A\mu} \right\rangle -
	\left\langle \mathcal E^A_\mu \mathcal E^{A\mu} \right\rangle,
\label{2-10}
\end{equation}
where $\mathcal E^A_\mu, \mathcal H^A_\mu$ are chromoelectric and chromomagnetic fields. We see that if $G(x) < 0$ then
the chromoelectric field is predominant in this area,
but if $G(x) > 0$ then the chromomagnetic field is predominant. Substitution of the ansatz \eqref{qcd-gl-140}-\eqref{qcd-gl-160} in the gluon condensate \eqref{2-10}
gives the following expression:
\begin{equation}
	G = - \frac{1}{2} {f'}^2 + \frac{1}{2} {v'}^2 - \frac{1}{2} f^2 v^2 +
	\frac{1}{2} m_1^2 f^2 - \frac{1}{2} m_2^2 v^2
	+ \frac{1}{2} {\phi'}^2 - \frac{1}{2} \left( f^2 - v^2 \right) \phi^2 +
	\frac{\lambda}{4} \left(
		\phi^2 - \mu^2
	\right)^2 .
\label{2-15}
\end{equation}
The profile of $G(x)$ is presented in Fig.~\ref{fig3}. Fig.~\ref{fig4} shows a schematical presentation of the flux tube with the distribution of the chromoelectric and chromomagnetic fields. We see that the flux tube has a core (in \textcolor{blue}{blue}) where the quantum fluctuations of the chromoelectric field and the longitudinal classical chromoelectric field $E^3_z$ are concentrated. These chromoelectric fields are confined by a belt (in \textcolor{red}{red}) filled with the chromomagnetic field. That is exactly what the dual QCD model says us:
In the dual superconductor picture of the QCD vacuum, chromomagnetic monopoles (creating a chromomagnetic field) condense into dual Cooper pairs, causing chromoelectric flux to be squeezed into tubes.

\subsection{Vacuum balls}

In classical field theories there exist regular solutions with finite energy and action: monopoles \cite{Hooft,Polyakov}, instantons, and so on. It is natural that the fields in such configurations are not zero everywhere. In this regard the following question makes a physical sense: whether do exist in quantum field theories somewhat similar solutions, but where a quantum average from all fields are equal to zero while the dispersion of the fields are nonzero? If such objects would exist, it would be very strange field configurations: it would be possible to tell that it is the objects created from vacuum as quantum averages from all fields are equal to zero, but quantum averages from dispersions of the fields are not equal to zero.

It is possible that such object is a glueball as it has no quarks, so there are no sources for gluon fields and therefore it is possible to assume that the gluon field in it has a zero expectation value but a nonzero expectation value of squares of the fields.

It is especially interesting if such objects would exist in the gravity interacting with some fields, and special interest would be represented by a situation when such object does not exist without gravitational forces. For example, such object filled with a quantum gravitating electromagnetic field can be a hypothesized Wheeler's geon.

\subsection{Nonperturbative vacuum}

In P quantum field theories vacuum is defined using annihilation operator $\hat a$ as
\begin{equation}
  \left. \left.
    \hat a \right| 0
  \right\rangle = 0.
\label{6d-10}
\end{equation}
This definition explicitly uses the notion of quantum and consequently can not be used for the definition of NP vacuum. Physically the difference between P and NP vacuums is in the following: the P vacuum is a sea of virtual quanta that appear and annihilate everywhere and always; the NP vacuum is a sea of fluctuating fields that fluctuate around a zero expectation value and the sea in a NP case is not a cloud of quanta. NP vacuum can have unusual properties. In Ref.~\cite{Dzhunushaliev:2012zb} possible properties of NP vacuum are discussed and a scalar field with LOG-potential for an effective description of NP vacuum is introduced.

One well known bad property of a P vacuum is that it has infinite energy as a consequence of the presence of zero-point fluctuations. This property leads to an incorrect value of the $\Lambda$-term in gravity.

\subsection{String-like theories}

NP quantization can give us an interesting interpretation of string theory. Let us imagine that there is a quantum field theory where two- and three-point Green's functions are zero. Nonzero point Green's functions are  $G_4, G_8, G_8, G_{16}, \cdots$ only. Let us imagine that the simplest Green's function $G_4$ is a sheet. Other Green's functions $G_8, G_8, G_{16}, \cdots$ are obtained from a 4-point Green's function
just as
in the Feynman diagram technique where all Green's functions are obtained from a 2-point Green's function (propagator). Physical interpretation of such a situation is that there is a string which swept a sheet which is the 4-point Green's function and all nonzero Green's functions can be interpreted as the interaction of strings.

In the path integral language the quantization is the Fourier transformation
\begin{equation}
  Z[J^A] = \int D \phi_A e^{i \left(
    S\left[\phi_A\right] + J^A \phi_A
  \right)}.
\label{string-10}
\end{equation}
Here $Z[J^A]$ is the generating functional; $S\left[\phi_A\right]$ is the action for the field $\phi_A$; $A$ is an generalized index; $J^A$ is the source field. The inverse Fourier transformation gives us the classical action
\begin{equation}
  S\left[ \phi_A \right] = \int D J^A e^{i \left(
    Z\left[J^A\right] + J^A \phi_A
  \right)}.
\label{string-20}
\end{equation}
Funny, but at such interpretation the quantization of ``string theory'' (which is described by ``the action'' $Z\left[J^A\right]$) gives us a quantum field theory with ``the generating functional'' $S\left[ \phi_A \right]$.

Thus: (a) in such interpretation a \textcolor{blue}{\emph{string is a quantum object described by a 4-point Green's function}} \cite{Dzhunushaliev:1996gb};
(b) quantization of quantum object brings us back to the original classical theory. In this situation, it is not surprising
that the initial classical theory has unusual properties. In particular, it exists only in a multidimensional space.

\section{Discussion and conclusions}

We have considered a NP quantization \`{a} la Heisenberg \cite{heis}. Calculations within the NP quantization are much more difficult than those within
a P quantization scheme. The truth is that in the first case we have to find physical arguments to cut off an infinite set of equations,
but in the second case only one problem exists -- the renormalization of loops. This is a usual situation: linear problems are always solved much more simpler than the nonlinear ones.
This can be demonstrated schematically as follows:
\begin{equation}
	\frac{\text{complexity of a \emph{nonperturbative}
	quantum theory}}
	{\text{complexity of a \emph{perturbative} quantum theory}}
	\gg
	\frac{\text{complexity of a \emph{nonlinear}
	classical theory}}
	{\text{complexity of a \emph{linear} classical theory}}
	\gg 1.
\label{concl-10}
\end{equation}
NP quantization procedure can be schematically written as
\begin{equation}
	\begin{matrix}
  \text{operator field equation/s} \\
  \downarrow \\
  \text{infinite set of equations for all Green's functions}\\
  \downarrow \\
  \text{cutting off (closure procedure)}\\
  \downarrow \\
  \text{solving of a finite system of equations}
\end{matrix}
\label{concl-20}
\end{equation}
Let us briefly list the main features of NP quantization:
\begin{itemize}
  \item NP quantization can in principle be applied for any theory. It does not depend on the renormalizability of given field theory: \textcolor{blue}{\emph{non-renormalizability does not mean non-quantizability.}}
  \item Green's functions are much softer on the light cone. The reason is that NP theories may have regular compact objects. In such objects quantum correlations between space separated points should be nonzero.
  \item There are common mathematical features between a NP quantization and turbulence modeling: in both cases, we have an infinite system  of equations for Green's functions/cumulants.
  \item For solving an infinite set of equations for all Green's functions the use of some physical arguments is absolutely necessary to cut off the infinite set of equations to a finite system of equations.
  \item \emph{Closure problem}:
  appearance of some problems associated  with the cutoff procedure is not excluded.
  \item The properties of field operators and quantum state of a given physical system are completely determined by all Green's functions.
\end{itemize}

\section*{Acknowledgements}

This work was supported by the Volkswagen Stiftung and by a grant in fundamental research in natural sciences by the Ministry of Education and Science of Kazakhstan. I am very grateful to V. Folomeev for fruitful discussions and comments.

\end{document}